# A Novel Approach for Periodic Assessment of Business Process Interoperability

**Badr Elmir, Bouchaib Bounabat**

**Al-Qualsadi Research & Development Team,**
**Ecole Nationale Supérieure d'Informatique et d'Analyses des Systèmes, ENSIAS,**
**Université Mohammed V Souissi,**
**BP 713, Agdal Rabat, Maroc**

**Abstract**
Business collaboration networks provide collaborative organizations a favorable context for automated business process interoperability. This paper aims to present a novel approach for assessing interoperability of process driven services by considering the three main aspects of interoperation: potentiality, compatibility and operational performance. It presents also a software tool that supports the proposed assessment method. In addition to its capacity to track and control the evolution of interoperation degree in time, the proposed tool measures the required effort to reach a planned degree of interoperability. Public accounting of financial authority is given as an illustrative case study of interoperability monitoring in public collaboration network.

## 1. Introduction

Organisations should prepare themselves to provide fully integrated services for their customers, partners and suppliers. In this context, horizontal cooperation within business collaborative networks has become a key enabler for e-business. Indeed, the delivery of value added online services often requires cooperation between two or more entities. This cooperation goes from simple information exchange and can reach business processes interoperability among collaborative businesses [1].
Establishing business process interoperability is not an end in itself, but an enabling capability for the strategic goal of collaboration establishment [2]. In this context, the present work focuses on monitoring the interoperability degree between automated business-processes involved in the provision of an integrated e-service. The studied processes may be located within a single organization or across organizational boundaries. In this sense, the proposed approach based on a measurement method [3] consists of five steps and takes into account the main aspects of interoperability.
The objective of this work is to: (1) Identify the most important characteristics of interoperability of business process driven services. This work proposes a set of criteria used to assess interoperability in this context considering all aspects of collaboration situation. (2) Describe a periodic assessment approach of interoperation driven by enterprise architecture paradigm. This allows knowing what is needed to reach a planned level of interoperability.
In this article, the second section is devoted to e-business system interoperability. The third section presents the "RatIop" assessment method that is based on a set of IT indicators for interoperability measurement. The fourth section proposes the monitoring approach model adopted in this study. It presents also the platform developed to support interoperability monitoring of business process driven services.

## 2. Interoperability in business collaboration networks

### 2.1 Interoperability

Interoperability characterizes the ability, for any number of processing information systems, to interact and exchange information and services between them. It requires a collective approach, an understanding of how each collaborating businesses operates and the development of arrangements which effectively manage business processes that cut across organizational boundaries [2].
Collaborative enterprises face technical and semantic difficulties [4] in order to establish interoperability. They face also organizational challenges [5]. Moreover, monitoring interoperability is not easy on such a macroscopic level.
In fact, interoperability is an information system quality that can be viewed from various perspectives. Several taxonomies have been proposed in this direction. In this sense, there are:

- Many levels of interoperability concern: business, process, service and data level [6].
- Various approaches to establish interoperability: integrated, federated, and unified approach [7].
- Multiple barriers could handicap interoperation: conceptual, organizational and technical barriers [8].
- Different scopes of application: within the same organization, cross independent organizations [9],
- Different transactional aspects of cooperation: synchronous or asynchronous collaboration [10].
- Diverse measurement perspectives: potentiality, compatibility, performance efficiency [11].

Thus, in terms of concern, there are various levels where interoperability takes place within a business collaboration network (BCN) [6]:

- Business level that refers to how to work within a business network in harmonized way in order to collaborate.
- Process level aims making various processes working together. In the case of a networked administration, internal processes of two entities are connected to create a common macro process.
- Service level is concerned with identifying, composing, and making function together with various applications.
- Data level refers to making synergy between different data models and heterogeneous conceptual schemas.

Also, in terms of barriers, the interoperability implementation faces [8]:

- Conceptual barriers which are related to the syntactic and semantic problems of information to be exchanged.
- Organizational barriers which refer to the definition of responsibilities and authority so that interoperability can take place.
- Technical barriers which deal with the use of adequate protocols, languages and infrastructure in communication.

In this case, interoperation compatibility check has to consider these barriers on each one of the four enterprise architecture layers cited below (Business, Process, Service, Data) [12].
Regarding the measurement aspects, [11] differentiate between the following complementary characteristics:

- Interoperation potentiality: it is an «internal quality» of a system that reflects its preparation to interoperate. This involves identifying a set of characteristics that have an impact on communication with partner's systems without necessarily having concrete information on them. The objective is to foster interoperability readiness by eliminating barriers that may obstruct the interaction.
- Interoperation compatibility: it r epresents an «external quality». In fact, interaction ability of two support systems is ensured through an engineering process aiming to establish interoperation between them.
- Interoperation performance: the third aspect characterizes the «quality in use». It focuses on monitoring operational performance. It consists of an assessment of the communication infrastructure availability, and the supporting system in general.

## 2.2 Service Oriented Business Collaboration

Collaboration in business context refers to the process where several organizations work together in an intersection of common goals [13], [14].
A business collaboration network (BCN) enables companies to communicate and collaborate with their customers, partners and suppliers in a productive way [15]. This cooperation takes different forms, from simple information exchange, to business processes interoperability among independent enterprises [16], [17].
Also, with the emergence of service provisioning environments, independent businesses become able to collaborate in order to have benefic results for all [18]. Among the main forms of cooperation, occurs integrated service providing to clients.
Hence, enterprises propose several online services in order to (a) improve their operations, (b) to make easy procedures (c) to accelerate revenues and (d) to minimize cost and time of services delivery [19]. Several works have listed most demanded e-services that business should provide [20], [21], and [22]: public information, online payment, e-procurement, e-commerce, value chain service, intermediation, etc.
In terms of service nature, online services have different use cases. Thus, authors of [23] differentiate between:

- Informational use: the service providers publish information to educate, entertain, influence, or reach their potential customers;
- Transactional use: enterprises support a coordinated sequence of system activities to provide services and transfer values;
- Operational use: when an organization provides a new mechanism to conduct business operations by integrating information systems into synergistic networks.

## 2.3 Interoperability of business process driven service

The concept of enterprise architecture (EA) attracted a lot of interest during the past decade [23]. It aims to provide a structure for business processes and systems that supports them. EA represents information systems using models in order to illustrate interrelationship between their components and the relationship with their ecosystems.

EA proposes to take an inventory of information system components by considering: (1) organization procedures, etc. (2) business process (3) IT applications, (4) technical infrastructure.

Indeed, most businesses around the world have established Enterprise Architecture programs [24]. They aim to eliminate overlapping projects, to support reuse, and to enhance interoperability.

On the other hand, several tactical plans were limited to the single issue of interoperation and many interoperability frameworks were developed. They mainly address technical problems by referencing the main recommended specifications to facilitate and promote cooperation within and between organizations [17].

In this sense and in order to facilitate interoperation within a business collaboration network, usually BCN members tend to adopt enterprise architecture as strategic choice of organization using "the service oriented" paradigm and techniques to implement and deploy services.

Service-oriented interaction model implements less coupled connections between various distributed software components. The approach seeks to provide abstraction by encapsulating functionality and allowing reuse of existing services [32].

In this case, An automated business process, as designed in Fig. 1, exposes to its clients a set of business services. This process may be elementary or composite. Composite processes are composed by a s et of processes. An elementary process ensures a set of activities. Theses automated activities use IT applications via application services.

An integrated business process may be located within a single organization or across organizational boundaries. In this context, clients expect to perceive business as a homogeneous and coherent unit in order to have a unified access to services they need. So, a BCN should be prepared to interact effectively with all the surrounding actors. This requires essentially openness and willingness to break functional, organizational and technological barriers.

A business process is a s et of related activities or operations which, together, create value and assist organizations to achieve their strategic objectives. A systematic focus on improving processes can therefore have a dramatic impact on the effective operation of agencies [2].

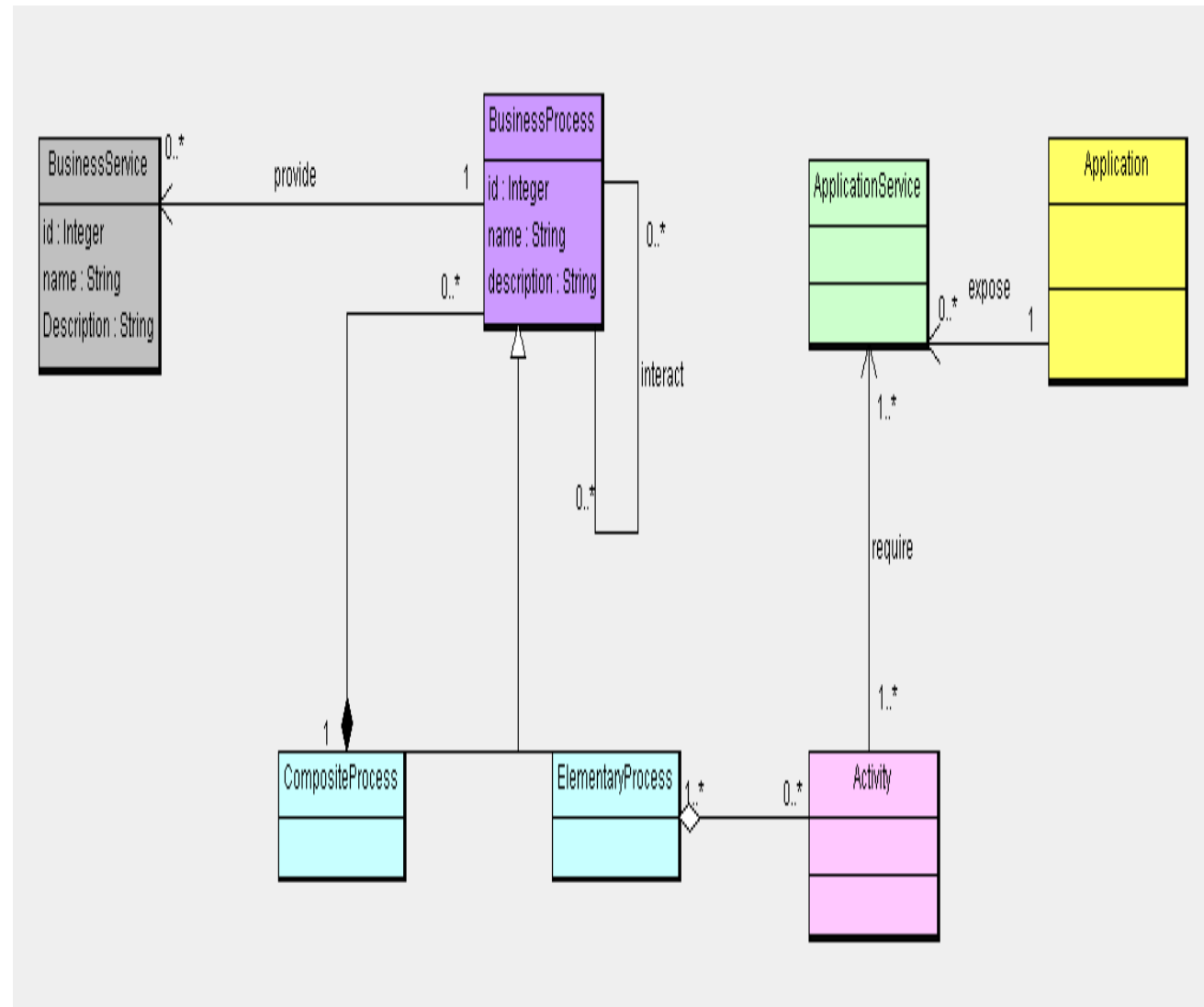

Fig. 1. Model of Business Process Driven Services

## 2.4 Interoperability indicators

Many works were interested on interoperability parameters that influence enterprise interoperability. Authors of [25] enumerate a set of factors influencing service interoperability and thereby of interest when performing analysis.

For instance, [26] captures a set of factors responsible for Business Interoperability in the context of Collaborative Business Processes. It proposes a q uotient measurement model for business interoperability.

On another hand, several interoperability maturity models (IMM) were introduced to describe the interoperation potentiality. They are mostly inspired by the CMM/CMMI model [27]. [28] Lists among others:

- ITIM (It Investment Management),
- LISI (Level of Information System Interoperability),
- OIMM (Organizational Interoperability Maturity Model),
- EIMM (Enterprise Interoperability Maturity Model),
- GIMM (Government Interoperability Maturity Matrix),
- SPICE (Software Process Improvement and Capability dEtermination).

Each model adopts a s pecific vocabulary to express the levels of maturity. However, the models have in general five scales ranging from low to high:

- An organization with a low level of interoperability is characterized as working independently or in isolation from other organizations and in an ad hoc or inconsistent manner.
- An organization with a high level of interoperability is characterized as being able to work with other organizations in a unified or enterprise way to maximize the benefits of collaboration across organizations.

In terms of compatibility and in order to dematerialize a business process and to interconnect it with its ecosystem, there is a necessity to study the external interfaces of its support systems. In this phase, the degree of compatibility «DC» is calculated on the basis of a mapping between the underlying components and the adjacent processes.

Several studies have focused on the characterization of the interoperation compatibility. For instance, [29] identifies several indicators to describe it.

The operational performance «PO» measurement is done on the basis of IT dashboards of involved organizations. It takes into account indicators as the availability score of the application servers, communication quality of service, and the end users degree of satisfaction about the interoperation in use. This information is collected based on surveying key end users.

Therefore, Interoperability assessment can be modeled as illustrated in Fig.2.

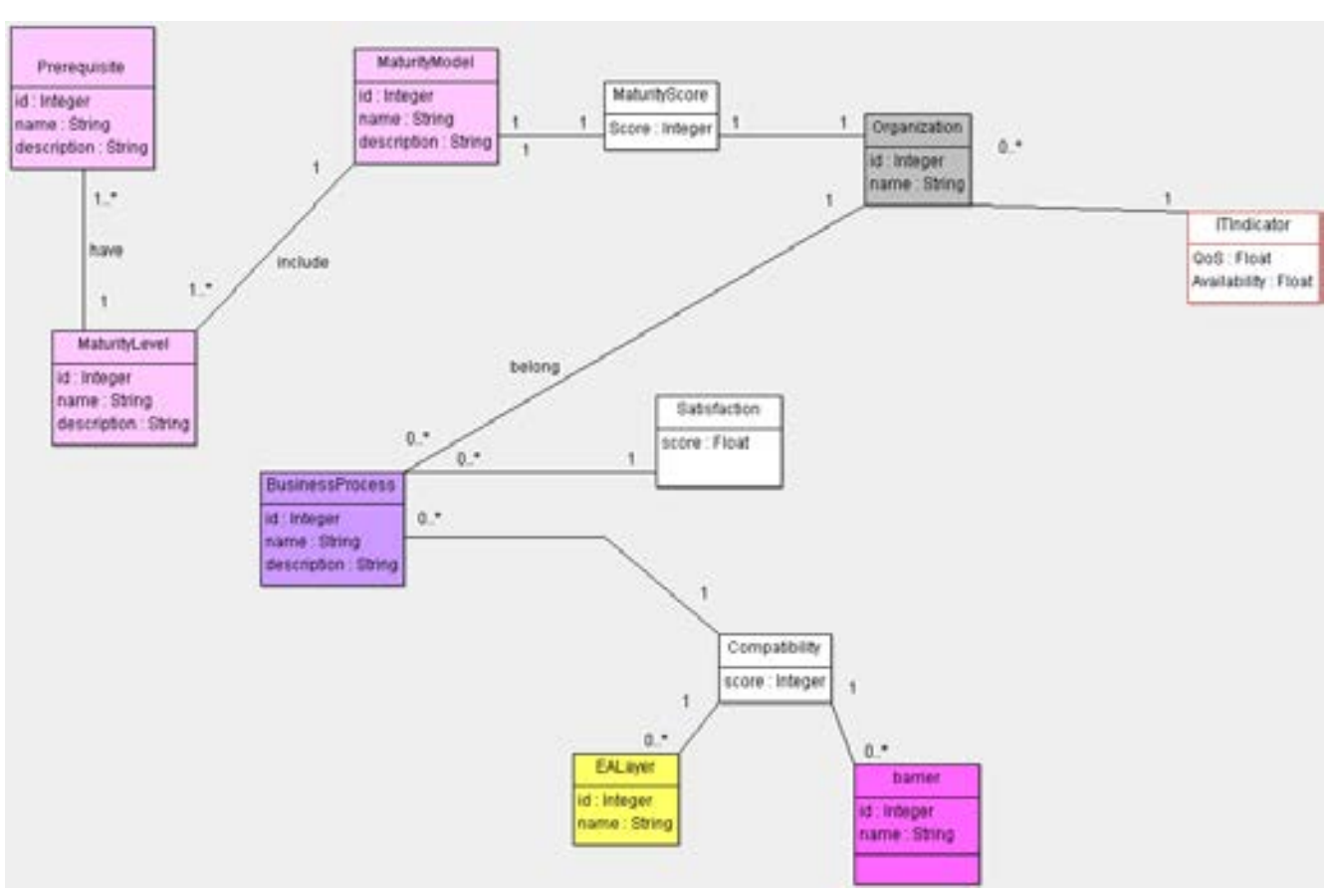

Fig. 2. Process interoperability assessment

## 3. Interoperability assessment

### 3.1 interoperability assessment method

The present section reminds the key elements of RatIop which is a five steps method that assess interoperability needed to deliver integrated business e-service [30]. These steps described in Fig. 3 are as follows:

1. Delineating the scope of the study.
2. Quantifying the interoperation potentiality.
3. Calculating the compatibility degree.
4. Evaluating the operating performance.
5. Aggregating the degree of interoperability.

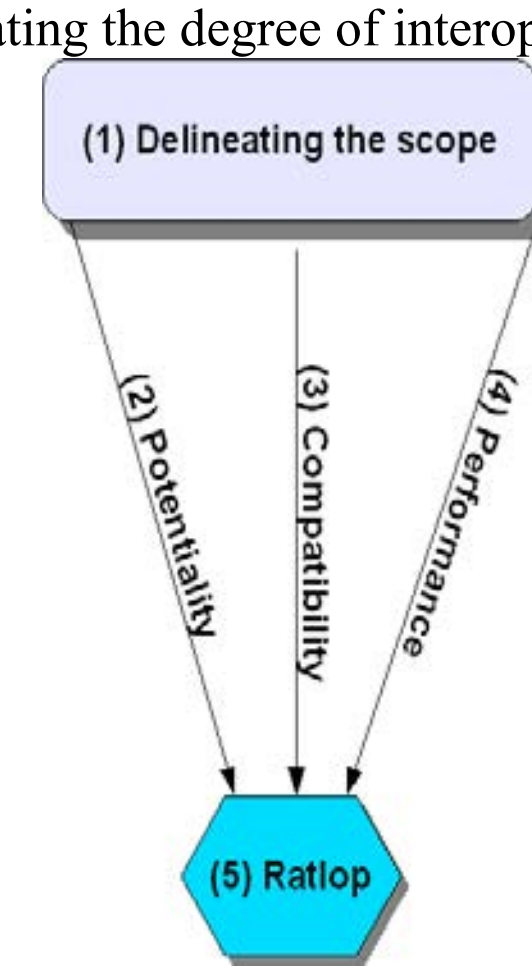

Fig. 3. Five steps of interoperability measurement [30]

### 3.2 Delineating the scope of the study

Assessing interoperability, whether used or required, to deliver business services requires the knowledge of its ecosystem.

In practical terms, the study focuses on a macro business process consisting of a s et of sub automated processes among independent business entities. These sub processes are linked together via several interfaces identified in advance. In this case, the preliminary phase consists of identifying the context of the studied automated business process then lists its underlying automated processes.

This step includes identifying:

- Organizations involved in the cooperation.
- Sub process within each entity in order to study compatibility.
- Information systems that support automated business processes within each organization in the BCN.
- Application services that enables sub processes interactions.

### 3.3 Quantifying the interoperation potentiality

The calculation of the potential for interoperability within the $k_{th}$ BCN member «$PI_k$» requires the adoption of one of these maturity models mentioned above. The organization is classified then on one of these five levels noted IMML (for interoperation maturity model level). To identify the potential degree of interoperability, we propose then the following mapping (See Table 1):

Table 1. Quantification of interoperation maturity

| Maturity Level (IMML) | Potentiality quantification |
|---|---|
| 1 | 0.2 |
| 2 | 0.4 |
| 3 | 0.6 |
| 4 | 0.8 |
| 5 | 1 |

Within each BCN member, the potential is calculated using the following equation

$$PI_k = 0.2 * IMML_k \quad (1)$$

The final interoperation potentiality is given by Equation 2 below:

$$PI = \min(PI_k) \quad (2)$$

### 3.4 Calculating the degree of compatibility

To assess the compatibility degree, the present work uses the compatibility matrix of Chen and Daclin [9].
The compatibility matrix, as presented in Table 2, consists of a combination of the "interoperability levels perspective" and "interoperability barriers perspective" seen in section 2.1. In practical terms, we enumerate conceptual, technical and organizational barriers in the different layers of interoperability concern: business, process, service and data.
By noting the elementary degree of interoperation compatibility «$dc_{ij}$» (i takes values from 1..4, and j takes values from 1..6).

Table 2. Interoperation compatibility

| | *Conceptual* | | *Organisational* | | *Technology* | |
|---|---|---|---|---|---|---|
| | syntactic | semantic | authorities responsibilities | Organisation | platform | communication |
| Business | $dc_{11}$ | $dc_{12}$ | $dc_{13}$ | $dc_{14}$ | $dc_{15}$ | $dc_{16}$ |
| Process | $dc_{21}$ | $dc_{22}$ | $dc_{23}$ | $dc_{24}$ | $dc_{25}$ | $dc_{26}$ |
| Service | $dc_{31}$ | $dc_{32}$ | $dc_{33}$ | $dc_{34}$ | $dc_{35}$ | $dc_{36}$ |
| Data | $dc_{41}$ | $dc_{42}$ | $dc_{43}$ | $dc_{44}$ | $dc_{45}$ | $dc_{46}$ |

Therefore, if the criteria in an area marked satisfaction the value 0 is assigned to $dc_{ij}$; otherwise if a lot of incompatibilities are met, the value 1 is assigned to $dc_{ij}$.
The degree of compatibility «DC» is given as follows:

$$DC = 1 - \sum (dcij / 24) \quad (3)$$

### 3.5 Evaluating operating performance.

By Denoting:

- «DS» the overall availability rate of application servers.
- «QoS» service quality of different networks used for interacting components communication. QoS is represented mainly by the overall availability of networks.
- «TS» end users satisfaction level about interoperation.

Given the cumulative nature of these three rates, the evaluation of operational performance is given by the geometric mean [31] as the following equation (See Equation 4):

$$PO = \sqrt[3]{(DS * QoS * TS)} \quad (4)$$

### 3.6 Aggregating the degree of interoperability

The final calculation of the ratio characterizing the interoperability - RatIop for ratio of Interoperability - process in question is by aggregating the three previous indicators using a function f defined in [0,1]3 → [0,1] (See Equation 5)

$$RatIop = f(PI, DC, PO) \quad (5)$$

Given the independent nature of these three indicators, we opt for the arithmetic mean [31] as follows (See Equation 6):

$$RatIop = (PI + DC + PO)/3 \quad (6)$$

In case the BCN has elements for pondering each one of these three indicators with different weights (w1, w2, w3); we choose the weighted arithmetic mean.

$$RatIop = (w1 * PI + w2 * DC + w3 * PO)/(w1 + w2 + w3) \quad (7)$$

## 4. Periodic Interoperability monitoring approach

### 4.1 Interoperability monitoring approach

It is quite interesting to analyze, track and control processes interoperability degree evolution from the existing "as-is" state to the future "to-be" state.
During each period, the proposed method supporting tool, (IMT) for interoperability monitoring tool, assesses interoperability. In addition to its capacity to track the evolution of interoperation degree in time, the IMT measures the required effort to reach a planned degree of interoperability.

(3)

This section shows the extended metamodel we propose for the interoperability monitoring approach. It includes all below aspects:

- The studied automated business processes.
- Connections and compositions that exist between the involved sub processes.
- Organizations member of the BCN participating in the business process interactions.
- Maturity model used in every organism, their levels and the prerequisites to reach each level.
- End users Satisfaction level.
- Enterprise architecture layers that coincides with the level of interoperation concerns.
- The elementary barriers that may obstruct interoperation situations.
- Periods within which interoperability is assessed.

The metamodel serves as a basis for interoperability assessment periodically. It considers existing IT indicators within the business collaboration network like availability rate of application servers and the network. It includes end users satisfaction about used interoperation. This metamodel includes maturity score of each involved organism. It references furthermore compatibility aspects on all levels of the enterprise architecture of collaboration in use.

The metamodel is represented by using UML diagrams in Fig. 4.

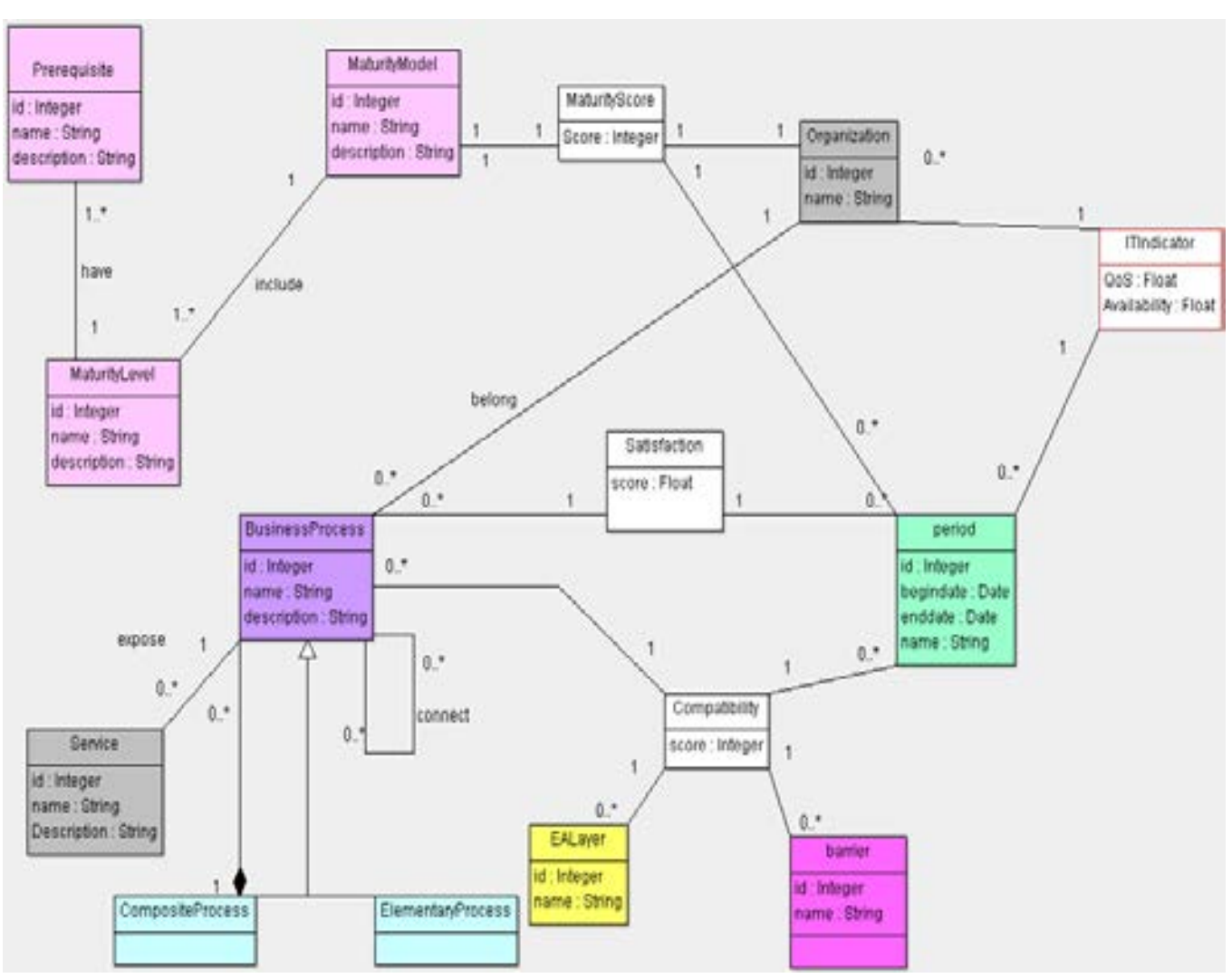

Fig. 4. Interoperability monitoring Metamodel (Class Diagram)

## 4.2 Periodic Interoperability monitoring tool

The Interoperability monitoring tool (IMT) includes three principal modules. The first one is dedicated to interoperability assessment at a s pecific period. Fig. 4 describes interoperability assessment of an automated macro process within private Business that uses EIMM as maturity model. In this specific case, we notice that there a lot of conceptual and organizational incompatibilities on a business layer.

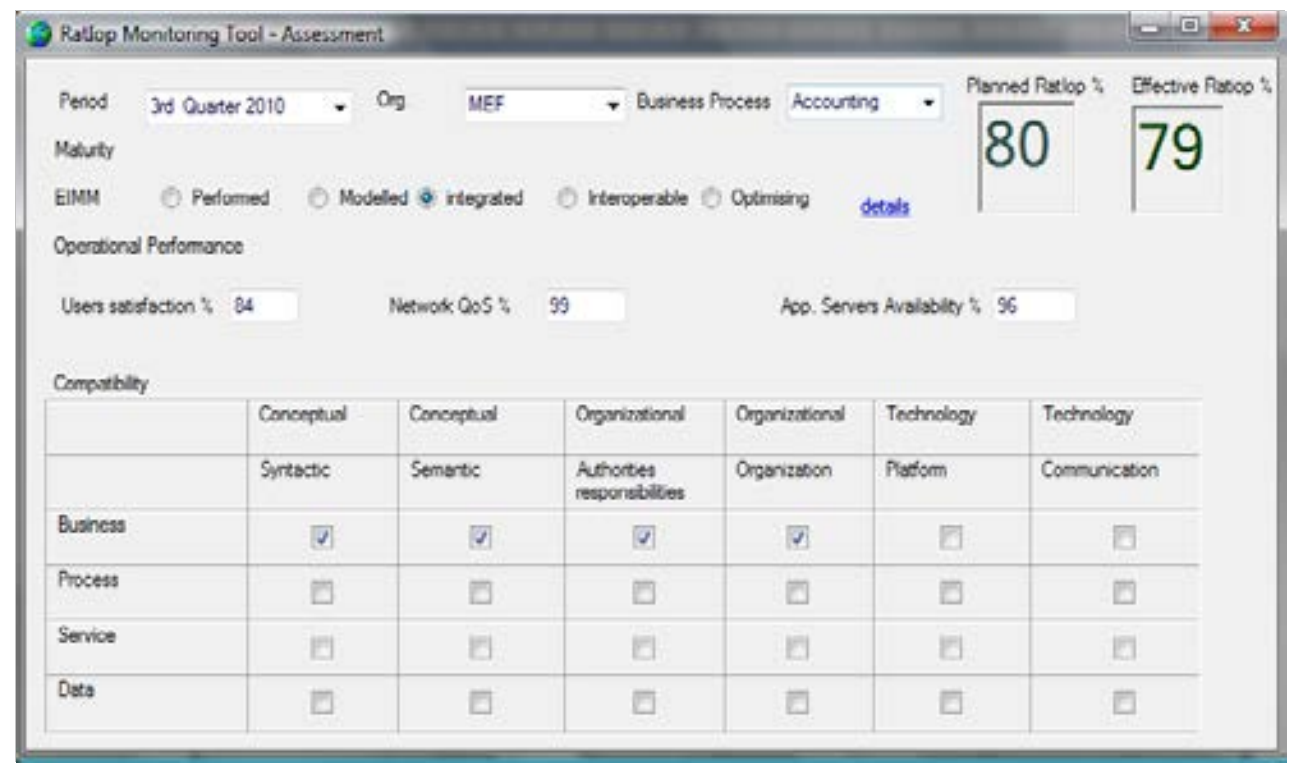

Fig. 5. Screen from Interoperability monitoring tool

In addition to its capacity to track the evolution of interoperation degree periodically, the IMT gives the possibility to propose a scenario to reach a planned degree of interoperability. For instance, in the example shown on Fig. 7, we plan to increase the interoperability ratio from an "As-is" RatIop to a "To-be" RatIop.

IMT proposes to (i) improve interoperability maturity to reach the fourth level, (ii) optimize the availability of involved application servers, (iii) better meet end users expectations and (iv) to resolve conceptual incompatibilities.

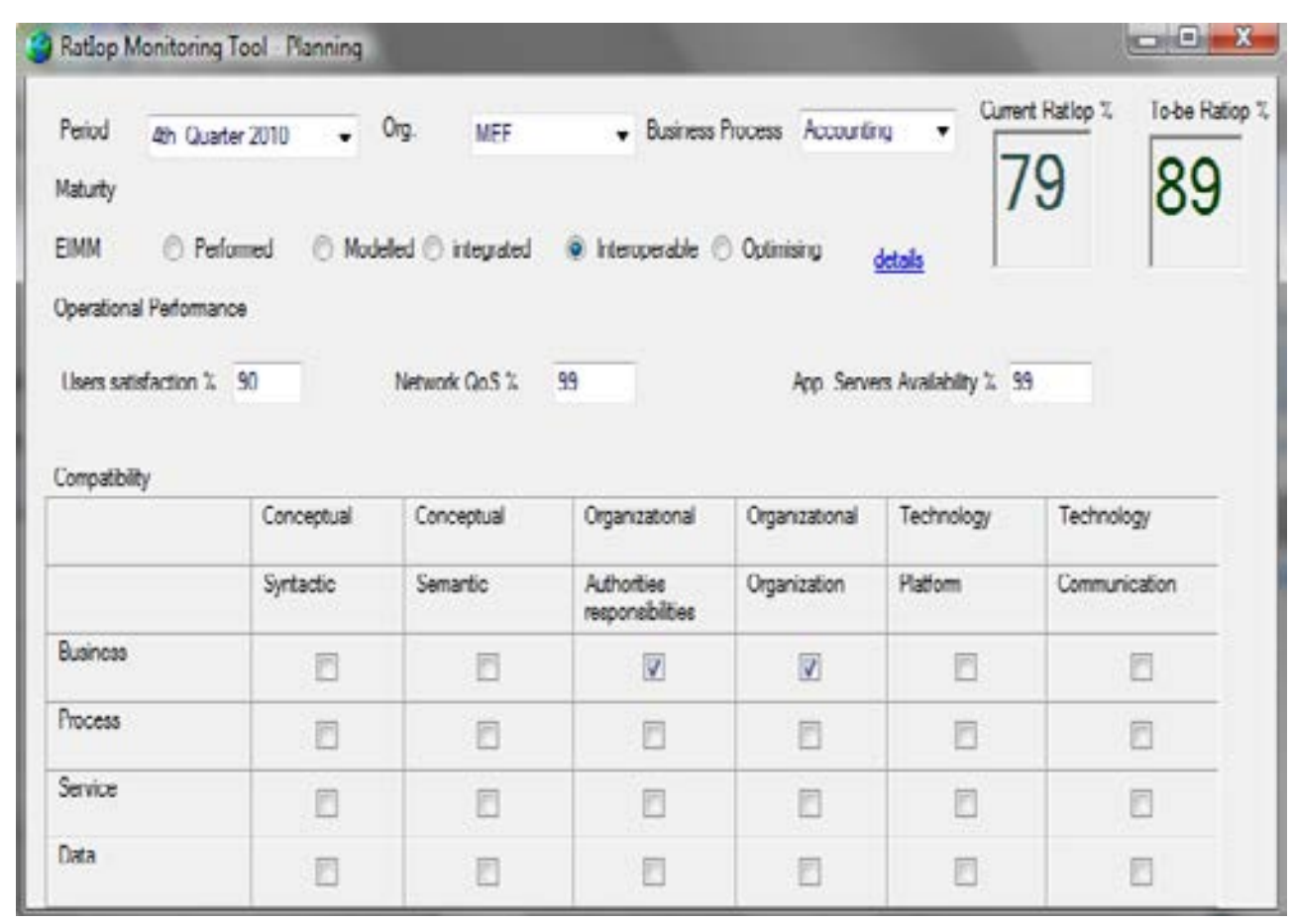

Fig. 6. Planning of Interoperability optimization

## 5. Case Study

This section is interested on the interoperability monitoring of Public Accounting Process. This process process is selected for (i) its interactions with other business processes within public Finance system (ii) and its willingness to deliver a set of integrated public services.
In general, Information System of the financial authority contains, among others, the following set of functional management subsystem:

1. Public Expenditure management system (S1)
2. Public Income management system (S2)
3. Public Accounting management system (S3)
4. Public Debt management (S4)

In our case, the three first systems (S1, S2 and S3) are located within the treasury department. The fourth system is managed by the Public Debt department.
S1 is linked with all authorizing officers within Ministries via an EDI System.
S2 is mainly connected via an ETL subsystem with:

- Integrated custom system (S5) managed by the custom administration
- Integrated Tax system (S6) Governed by the Tax department.

S5 and S6 offer transactional services to citizens and Private sector.

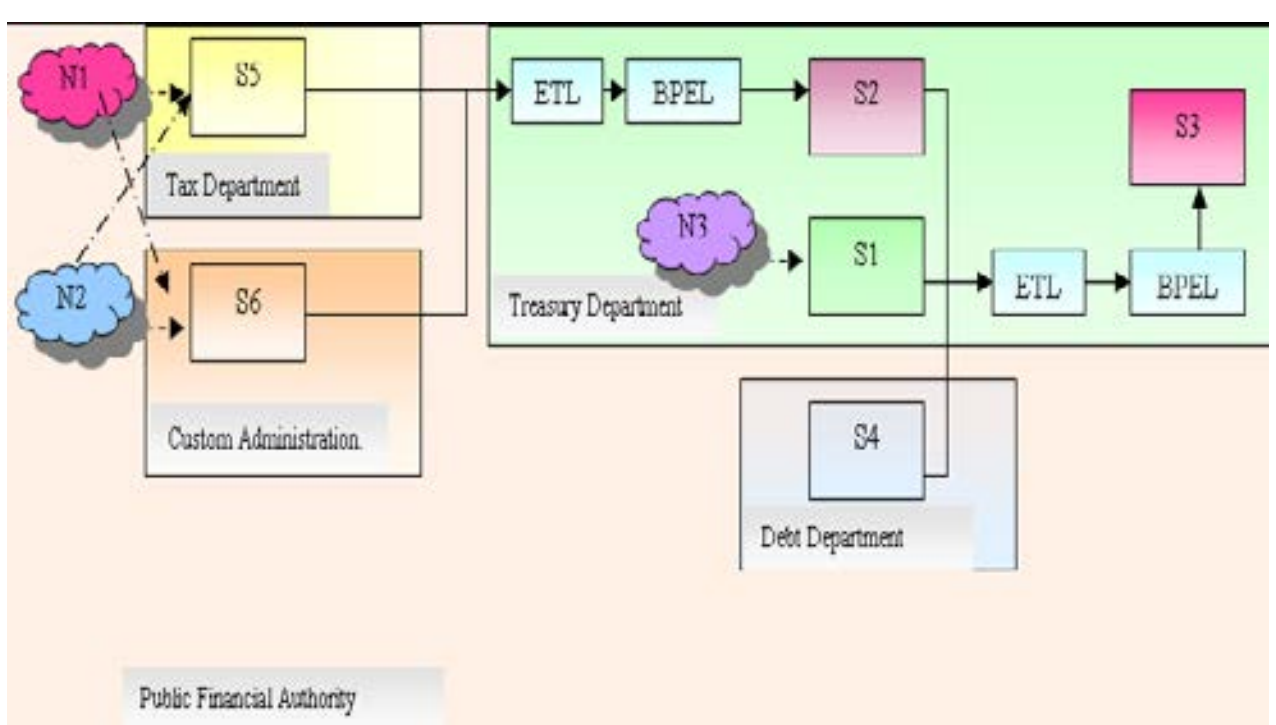

Fig. 7. Extract of a Public Financial Authority process interaction

N1 refers to citizens, N2 refers to the Private sector and N3 refers to the Authorizing officers within ministries.
If we focus on the accounting process automated with S3, its interactions with debt, income and expenditure processes occur via an ETL (Extract, Transform, Load) tool connected with specific BPEL processes (Business process execution language).
The interoperability is assessed every quarter in this collaboration network. During this period, the maturity level is brought up. The end user satisfaction is improved. The IT indicators are optimized (See Figure 8).

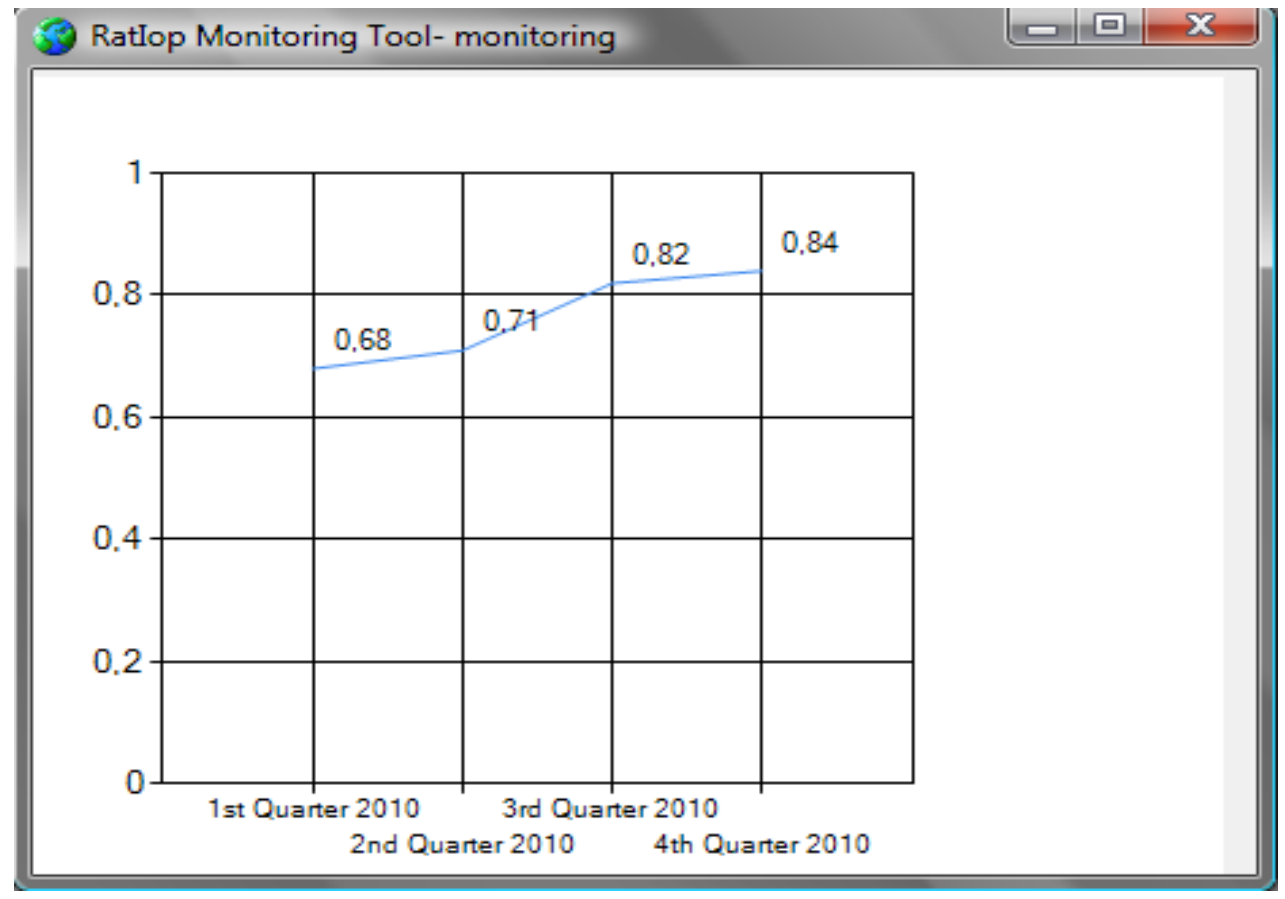

Fig. 8. Quarterly Interoperability monitoring of public Accounting

## 6. Discussion

The provided case study illustrates the results of interoperability monitoring within a specific collaboration context. This is provided by the interoperability monitoring tool (IMT) that collects existing IT indicators like availability score of Application servers and the end users satisfaction level. The IMT logs, in a convenient way, the interoperation incompatibilities in every enterprise architecture layer surrounding the accounting process.
This automated method enables the establishment of an action plan that aims to improve inter system integration.
The IMT is able to propose a scenario to reach a planned result for interoperability ratio. However the current version of the IMT is not yet able to propose the best scenario to achieve this result efficiently. This work is a prerequisite for several projects launched in parallel, and dealing with the applicability of control theory in the information system interoperability field. In this case, the future version of the IMT will use such techniques to enable the optimal control of interoperability.

## 7. Conclusions

Business networks enable organizations to collaborate in order to reach their common goals. This is in general insured via inter-organizational business process interoperability. The present paper proposes a n ovel approach based on a five steps assessment method that uses existing indicators within involved BCN members like quality maturity indicators, information technology dashboards, etc.
The result of the assessment method is a ratio metric enabling the measurement of this quality by taking into account three main operational aspects: interoperation potentiality, interoperation compatibility and operational performance.
The proposed approach is supported by a software tool IMT (for interoperability monitoring tool) that assess, track and control interoperability in this context.

**B. ELMIR** is a Software Engineer graduated from ENSIAS (2002) (National Higher School for Computer Science and S ystem analysis), holder of an Extended Higher Studies Diploma from ENSIAS (2006) and “Ph.D. candidate” at ENSIAS. His research focuses on interoperability optimal control in public administration. He is an integration architect on the Ministry of Economy and Finance of the Kingdom of Morocco since 2002. He also oversees the IT operating activity in this department.

**B. BOUNABAT:** PhD in Computer Sciences. Professor in ENSIAS, (National Higher School for Computer Science and System analysis), Rabat, Morocco. Responsible of "Computer Engineering" Formation and Research Unit in ENSIAS, International Expert in ICT Strategies and E-Government to several international organizations, Member of the board of Internet Society - Moroccan Chapter.